\documentclass[aps,prl,twocolumn,superscriptaddress,showpacs]{revtex4-1}
\usepackage{graphicx}
\usepackage{hyperref}
\usepackage{ucs}
\usepackage{amssymb}
\usepackage[colorinlistoftodos]{todonotes}
\usepackage{textcomp}

\begin{document}

\title{Spin excitations in a single La$_2$CuO$_4$ layer}

\author{M. P. M. Dean}
\email{mdean@bnl.gov}
\affiliation{Department of Condensed Matter Physics and Materials Science,
Brookhaven National Laboratory, Upton, New York 11973, USA}

\author{R. S. Springell}
\affiliation{London Centre for Nanotechnology and Department of Physics and Astronomy, University College London, London WC1E 6BT, United Kingdom}

\author{C. Monney}
\author{K. J. Zhou}
\affiliation{Swiss Light Source, Paul Scherrer Institut, CH-5232 Villigen PSI, Switzerland}

\author{I. Bo\v{z}ovi\'{c}}
\affiliation{Department of Condensed Matter Physics and Materials Science, Brookhaven National Laboratory, Upton, New York 11973, USA}

\author{J. Pereiro}
\affiliation{Department of Condensed Matter Physics and Materials Science, Brookhaven National Laboratory, Upton, New York 11973, USA}
\affiliation{Division of Physics and Applied Physics, Nanyang Technological University,
Singapore 6373616, Singapore}

\author{B. Dalla Piazza}
\affiliation{Laboratory for Quantum Magnetism, \'E{}cole Polytechnique F\'{e}d\'{e}rale de Lausanne (EPFL), CH-1015, Switzerland}

\author{H. M. R\o{}nnow}
\affiliation{Laboratory for Quantum Magnetism, \'E{}cole Polytechnique F\'{e}d\'{e}rale de Lausanne (EPFL), CH-1015, Switzerland}

\author{E. Morenzoni}
\affiliation{Laboratory for Muon Spin Spectroscopy, Paul Scherrer Institut, 5232 Villigen PSI, Switzerland}

\author{J. van den Brink}
\affiliation{Institute for Theoretical Solid State Physics, IFW Dresden, D01171 Dresden, Germany}

\author{T. Schmitt}
\affiliation{Swiss Light Source, Paul Scherrer Institut, CH-5232 Villigen PSI, Switzerland}

\author{J. P. Hill}
\email{hill@bnl.gov}
\affiliation{Department of Condensed Matter Physics and Materials Science, Brookhaven National Laboratory, Upton, New York 11973, USA}

\newcommand{\microns}{\ensuremath{\mathrm{\mu}}m}
\newcommand{\e}[1]{$\times$10$^{#1}$}
\def\mathbi#1{\ensuremath{\textbf{\em #1}}}
\newcommand{\pipi}{($\frac{\pi}{2}$,$\frac{\pi}{2}$)}
\newcommand{\LCO}{La$_2$CuO$_4$}

\date{\today}

\maketitle

\textbf{
The dynamics of $S=\frac{1}{2}$ quantum spins on a 2D square lattice lie at the heart of the mystery of the cuprates \cite{Hayden2004,Vignolle2007,Li2010,LeTacon2011,Coldea2001,Headings2010,Braicovich2010}.
In bulk cuprates such as \LCO{}, the presence of a weak interlayer coupling stabilizes 3D N\'{e}el order up to high temperatures. In a truly 2D system however, thermal spin fluctuations melt long range order at any finite temperature \cite{Mermin1966}. Further, quantum spin fluctuations transfer magnetic spectral weight out of a well-defined magnon excitation into a magnetic continuum, the nature of which remains controversial \cite{Sandvik2001,Ho2001,Christensen2007,Headings2010}. Here, we measure the spin response of \emph{isolated one-unit-cell thick layers} of \LCO{}. We show that coherent magnons persist even in a single layer of \LCO{} despite the loss of magnetic order, with no evidence for resonating valence bond (RVB)-like spin correlations \cite{Anderson1987,Hsu1990,Christensen2007}. Thus these excitations are well described by linear spin wave theory (LSWT).  We also observe a high-energy magnetic continuum in the isotropic magnetic response. This high-energy continuum is not well described by 2 magnon LSWT, or indeed any existing theories.
}

\begin{figure}
\includegraphics{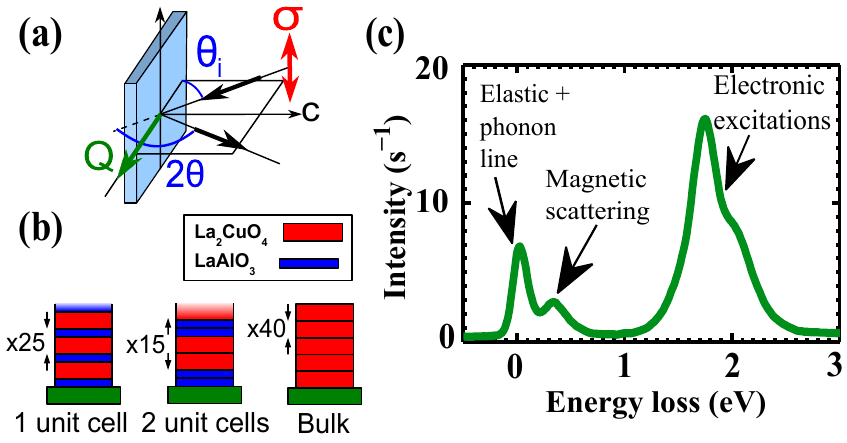}
\caption{(a) The experimental scattering geometry. The 931~eV $\sigma$ polarized x-rays are incident at an angle $\theta_i$ and are scattered through a fixed angle $2\theta= 130^{\circ}$.  Large $\mathbi{Q}$ corresponds to near-grazing incidence ($\theta_i \rightarrow 0$). (b) The multilayer films studied, composed of 13.2~\AA{} \LCO{} layers (red blocks) and 3.8~\AA{} LaAlO$_3$ (blue blocks). We label the films, based on the thickness of \LCO{}, as bulk, 2 unit cell and 1 unit cell. The arrows denote the repeat unit of the films ($\times40$, $\times15$, $\times25$). (c) A representative RIXS spectrum of the 1 unit cell \LCO{} film at \mathbi{Q}= (0.77$\pi$,0) identifying the main spectral features: the elastic and phonon scattering around zero energy transfer, the magnetic scattering around 300~meV and the electronic ($dd$) excitations from 1 to 3~eV.  \label{Figintro}}
\end{figure}

The simplest model for describing the magnetic excitations of undoped cuprates is LSWT \cite{Manousakis1991}. Coherent transverse magnetic excitations correspond to spin waves -- magnons -- with a well-defined energy; whereas longitudinal magnetic excitations result in a high-energy continuum of multi-magnons. While measurements of the long wavelength magnetic excitations of \LCO{} \cite{Keimer1992} can be understood in terms of a renormalzied classical model \cite{Chakravarty1989}, the short range correlations remain controversial  \cite{Sandvik2001,Christensen2007,Headings2010,Suter2011,Anderson1987, Ho2001} as quantum fluctuations can transfer spectral weight out of the magnon peak into a high-energy continuum.  Furthermore, the magnetic excitation spectrum of a truely 2D single \LCO{} layer, where spin fluctuations are strongest, has not been measured. This is because most of what we know about the spin excitation spectrum of the cuprates has come from inelastic neutron scattering \cite{Tranquada2005}. Unfortunately, such experiments require large samples and are often challenging at high energy transfers. In recent years, however, resonant inelastic x-ray scattering (RIXS) has achieved suffcient resolution to access magnetic excitations \cite{Braicovich2010,Schlappa2009,Guarise2010,Ament2011} and RIXS is well suited to measuring high-energy magnetic excitations in the range 100-1000~meV. Furthermore, the high sensitivity of the technique allows us to look at nanostructured samples and this in turn opens up the exciting possibility of measuring the spin response of a single \LCO{} layer for the first time.

We performed RIXS measurements on bulk and single layer \LCO{} films at 15~K using the scattering geometry shown in Fig.\ \ref{Figintro}(a).  The sample was rotated about the vertical axis to vary \mathbi{Q}, the projection of the total scattering vector in the $ab$-plane. To provide sufficient scattering volume of isolated \LCO{} layers, we prepared heterostructures based on 1 unit cell (uc) thick layers of \LCO{} and LaAlO$_3$. The samples are depicted in Fig.\ \ref{Figintro}(b) as 1uc =[1uc \LCO{} + 1uc LaAlO$_3$]$\times$25, 2uc =[2uc \LCO{} + 2uc LaAlO$_3$]$\times$15, and bulk =[1uc \LCO{}]$\times$40.

The RIXS spectra of the three samples were measured from ($0.14\pi$,0) to ($0.8\pi$,0) and ($0.1\pi$,$0.1\pi$) to ($0.6\pi$,$0.6\pi$). Figure \ref{Figintro}(c) plots a representative spectrum collected at \mathbi{Q}= (0.77$\pi$,0). We observe a peak corresponding to elastic scattering around zero energy transfer, which also contains a shoulder at low energy transfers (up to $\sim$90~meV) due to phonon scattering. From 200-800~meV we observe magnetic scattering and in the 1-3~eV window we identify electronic $dd$-excitations.

\begin{figure}[t]
\includegraphics{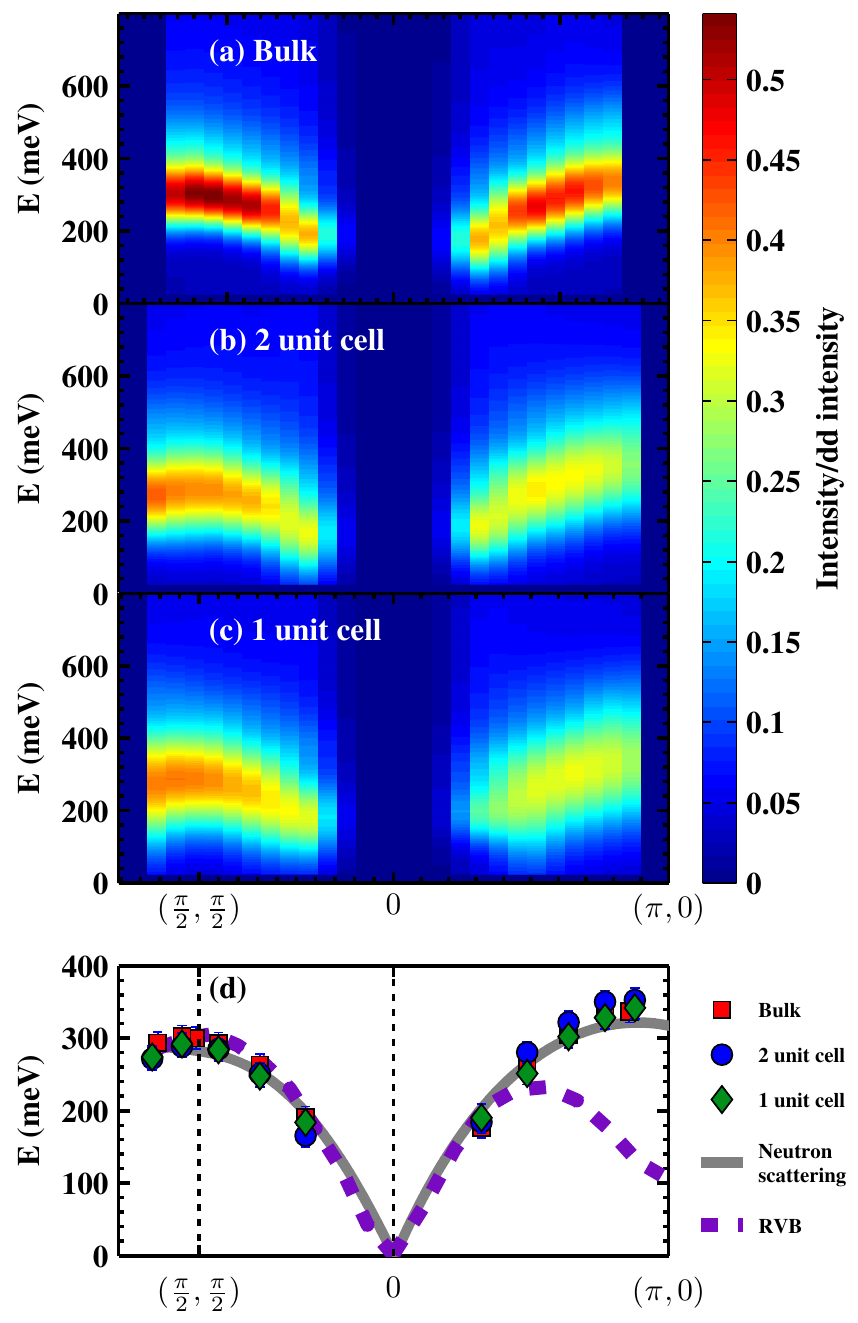}  
\caption{(a)-(c) The magnetic RIXS scattering intensity along the high symmetry lines in the BZ for: (a) Bulk film, (b) 2 unit cell and (c) 1 unit cell samples. (d) The peak energy dispersion in bulk (red $\square$ ), 2 unit cell (blue $\circ$) and 1 unit cell \LCO{} (green $\diamond$ ). The solid gray line is the result from neutron scattering measurements of bulk \LCO{} \cite{Headings2010}; the dotted purple line are calculations for an RVB model \cite{Hsu1990}.  \label{Figcolourmap}}
\end{figure}

The elastic and phonon peak in the spectra were fit using Gaussian functions and subtracted from the data, in order to isolate the magnetic scattering. In the bulk film the response is dominated by a dispersing magnon peak (see Fig.\ \ref{Figcolourmap}(a)), along with additional scattering extending out to higher energies. As $\mathbi{Q} \rightarrow 0$ the specular reflection from the sample surface overwhelms any magnetic signal. For the case of the 1uc and 2uc films, shown in Fig.\ \ref{Figcolourmap}(b) and (c), the peak intensity is suppressed and the peak width is broadened with additional weight at high energies.

Figure \ref{Figcolourmap}(d) plots the dispersion of the peak in the magnetic response of the three films. We see that the 1uc and 2uc films still display a coherent magnon peak, with the same dispersion as in the bulk. For comparison we also plot neutron scattering results from bulk \LCO{} \cite{Headings2010}. These are in excellent agreement along (0,0) to \pipi{}. Along (0,0) to ($\pi$,0) our results appear slightly higher in energy than Ref.\ \cite{Headings2010}. We attribute this to the fact that in the present case we are recording the median energy of the asymmetric peak, rather than the peak energy position, as is the case in Ref.\ \cite{Headings2010}. This difference has a bigger effect along the ($\pi$,0) direction because of the increased high-energy tail in this direction, a fact we will return to later.

Figure \ref{Figcolourmap}(d) shows the principal result of this paper: Even in \LCO{} layers only a single unit cell thick, a coherent, bulk-like magnon is a reasonable description of the magnetic excitations. Significantly, the dispersion is very similar to that of bulk \LCO{}. Thus, even though LSWT is based on an ordered N\'{e}el state, it continues to provide a reasonable description of the spin response of a truly 2D \LCO{} layer with no N\'{e}el order.

This is an important result. Recent muon spin rotation data seem to indicate that thin \LCO{} layers are dominated by quantum fluctuations \cite{Suter2011}. Indeed, recent calculations suggest that quantum fluctuations may be enhanced in \LCO{} because of frustrated higher order hoping \cite{Piazza2011}. Therefore, a key unanswered question is whether a single layer of \LCO{} hosts a more RVB-like state, or whether it obeys the expectations of the Heisenberg model with renormalized classical correlations? RVB-like models predict a much lower energy at ($\pi$,0) than at \pipi{} \cite{Hsu1990}. In contrast, Fig.\ \ref{Figcolourmap}(d) shows no such downturn and further our data imply the presence of similar magnetic correlations in single layer and in bulk \LCO{}. Thus, our results strongly support the second scenario.

More specifically, our single layer data are consistent with the renormalized classical expectation for the Heisenberg model: Even though thermal fluctuations suppress long range order at any finite temperature in a true 2D system, as the temperature is lowered, the correlation length increases exponentially in $J/k_{B}T$ \cite{Chakravarty1989} the spin wave lifetime likewise increases as correlation length divided by spin wave velocity \cite{Ronnow2001}, and the system mimics a N\'{e}el state in its response. Neutron scattering studies of the nearest-neighbor material Cu(DCOO)$_2\cdot$4D$_2$O have shown a small quantum correction to the magnon energy at ($\pi$,0) \cite{Christensen2007,Singh1995}. In contrast our data show a 40~meV (13\%) higher energy at ($\pi$,0) than at \pipi{}, thereby confirming the neutron measurements on bulk \LCO{} \cite{Coldea2001,Headings2010}. We conclude that in the single \LCO{} layer, the subtle quantum dispersion is hidden by a larger zone boundary dispersion due to the longer ranged interactions resulting from higher order hopping terms \cite{Coldea2001,Headings2010,Guarise2010,Piazza2011}.

\begin{figure}
\includegraphics{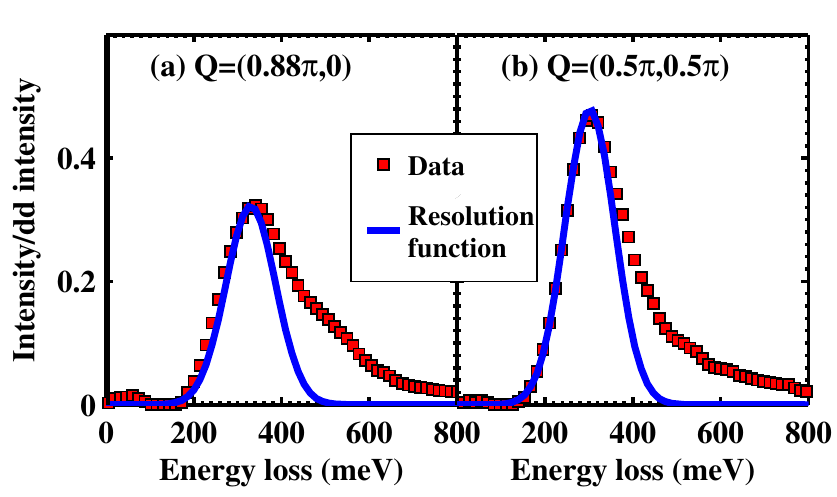}  
\caption{The measured magnetic spectral weight (red $\square$) in bulk \LCO{} at (a) \mathbi{Q}=(0.88$\pi$,0) and (b) \mathbi{Q}=(0.5$\pi$,0.5$\pi$). The blue line is a resolution-limited Gaussian at the single magnon energy.  \label{Fig_PoleCont}}
\end{figure}

We are now in a position to study the 2D $S=\frac{1}{2}$ response in detail. To do so, we first consider the bulk response, Fig.\ \ref{Fig_PoleCont}. A continuum of magnetic excitations is observed above the single magnon energy extending out to 800~meV, beyond which the tail of the $dd$ excitations makes it hard to determine the origin of the scattering. Similar high-energy continua are seen at all momenta, as is evident from Fig. \ref{Figcolourmap}. We note that neutron scattering  also sees high-energy magnetic scattering up to the maximum measured energy transfer of 450~meV \cite{Headings2010}. In Fig.\ \ref{Fig_PoleCont}, we see that this scattering is anisotropic, as this high-energy weight is stronger near ($\pi$,0) than at \pipi{} relative to the magnon peak.

In first-order LSWT, the magnons are fluctuations transverse ($T$) to the spin quantization axis, and result in a resolution-limited magnon peak, with no
weight at higher energies. Quantum fluctuations shift weight to a higher-energy continuum, which within SWT is longitudinal ($L$) two-magnon scattering. Quantum Monte Carlo calculations \cite{Sandvik2001} and neutron scattering \cite{Ronnow2001} for a nearest-neighbor Heisenberg antiferromagnet also indicated a transverse continuum especially around ($\pi$,0). It has been speculated that this can in part be described as factional spinon-like excitations \cite{Anderson1987, Ho2001}. Furthermore there is the prospect that RIXS can excite additional three-magnon processes \cite{Ament3mag,Igarashi2011}.

\begin{figure}
\includegraphics{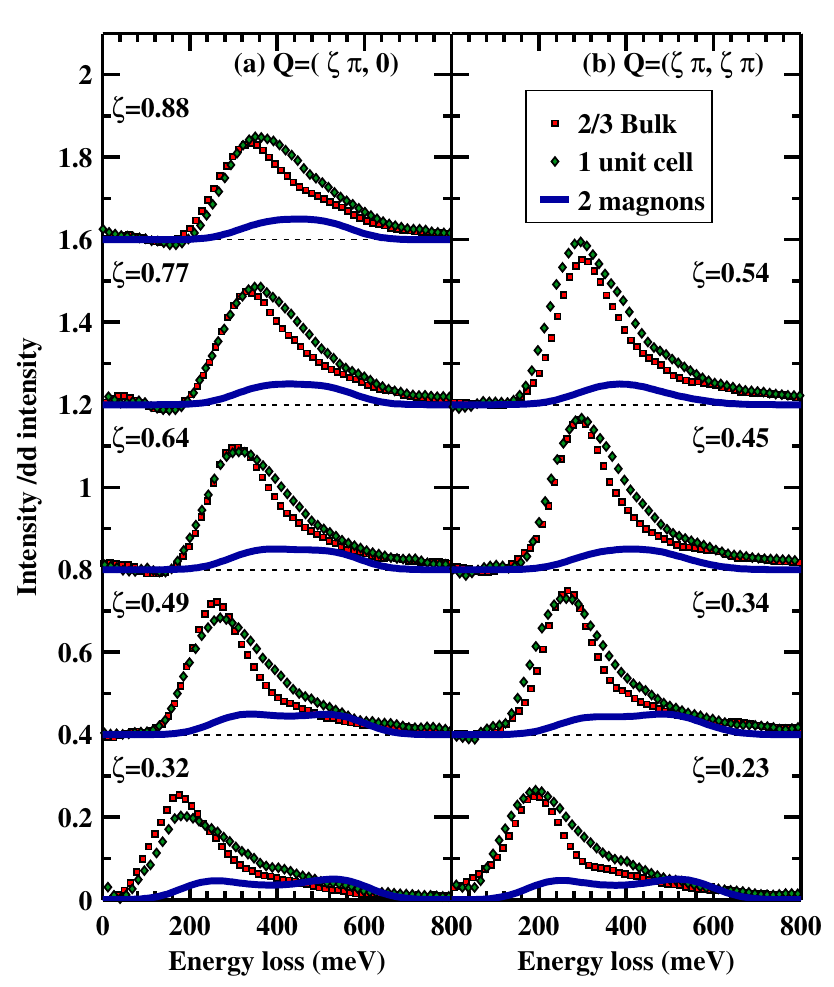}  
\caption{A comparison between $\frac{2}{3}$ of the bulk \LCO{} response (red $\square{}$) and the 1 unit cell response (green $\diamond$) along (a) ($\zeta \pi$, 0) and (b) ($\zeta \pi$, $\zeta \pi$). The difference between these spectra corresponds to the longitudinal magnetic response, to be compared to the two magnon  $S^{zz}(\mathbi{Q},\omega)$ calculation (blue line) \cite{Piazza2011}. As it is not currently  possible to measure absolute magnetic scattering intensities with RIXS, the calculation is shown with an arbitrary intensity.   \label{Fig_Sx_Sz}}
\end{figure}

With this description of the bulk scattering in hand, we now turn to the 1uc film for which there is no broken spin rotational symmetry. Comparing Fig.\ \ref{Figcolourmap}(a) and (c) we find that the 1uc film shows a smaller peak intensity than the bulk and more spectral weight at high energies. As shown in Refs.\ \cite{Ament2009,Haverkort2010,Igarashi2011}
Cu $L$-edge RIXS, with $\sigma$ polarized incident light, is dominated by the out-of-plane $c$-axis magnetic response. In the bulk sample, magnetic moments order in the plane along the (110) axis, which thereby defines the spin quantization axis.
Thus, here RIXS measures the transverse response $T$. However, in the 1uc film, the isolated \LCO{} layers  do not order \cite{Suter2011}. One then measures the isotropic response $\frac{2}{3}T+\frac{1}{3}L$. Figure \ref{Fig_Sx_Sz} compares the 1uc spectrum to $\frac{2}{3}$ of the bulk spectrum along (a) the ($\zeta \pi$, 0) and (b) the ($\zeta \pi$, $\zeta \pi$) symmetry directions.
The $\frac{2}{3}$ scaling factor results in similar heights of the single magnon peak, which is known to be transverse. This validates our approach and allows one to distinguish transverse and longitudinal components, leading to two important observations: i) A large component of the continuum is present in both data-sets, meaning that it is transverse in nature and therefore cannot be described as 2-magnon scattering in LSWT. ii) At high energy transfers the 1uc shows additional scattering, which is longitudinal in nature. Unlike the continuum response measured in the bulk film, we observe no clear differences between (0,0)$\rightarrow$($\pi$,0) and (0,0)$\rightarrow$\pipi{} for this additional scattering, which disperses to higher energies at higher \mathbi{Q}. In a LSWT picture, such additional longitudinal scattering would come from two-magnon $S^{zz}(\mathbi{Q},\omega)$ excitations. These are calculated for \LCO{} (see methods) and plotted for comparison in Fig.\ \ref{Fig_Sx_Sz}. We see that the two-magnon response does not accurately account for the additional scattering in the 1uc magnetic excitation spectra.

The ratio of \mathbi{Q}-integrated longitudinal to transverse scattering can be compared to calculations within LSWT \cite{Huberman2005}. For a Heisenberg magnet, this ratio is controlled by the spin reduction due to zero-point fluctuations $\Delta S = S - \langle S^z \rangle$, where  $\Delta S = 0.2$ for $S = \frac{1}{2}$ and then
\begin{equation}
\frac{\rm{Longitudinal}}{\rm{Transverse}}=\frac{ \Delta S (\Delta S + 1) }{ (S-\Delta S)(2\Delta S + 1) } \approx 0.6 .
\end{equation}
We find that the mean weight of the additional scattering relative to the transverse scattering at the measured \mathbi{Q} values is about 0.6, consistent with this estimate, though we note that our measured \mathbi{Q} does not constitute a full integration over the zone.

The comparision of bulk and 1uc \LCO{} have revealed a longitudinal and transverse continuum in the magnetic excitation spectrum of \LCO{}, which has yet to be fully explained. Possibilities include transverse incoherent scattering \cite{Sandvik2001}, higher-order magnons \cite{Ament3mag,Igarashi2011} and the presence of proposed \cite{Ho2001} spinon-like excitations in \LCO{} \cite{Headings2010}. Alongside this, the clearly defined magnon dispersion in 1uc LCO{} leads to the conclusion that even in the absence of long range order, the ground state hosts classical correlations rather than RVB-like quantum disorder. Further theoretical and experimental studies are called for to explain our observations, and to determine whether these non-LSWT features gain importance upon doping and whether they are relevant to the mechanism of superconductivity in doped cuprates.

\section{Acknowledgements}
We thank Robert Konik, Maurits Haverkort and Andrew Boothroyd for fruitful discussions, Xuerong Liu for assistance with the sample characterization and Stephen Hayden and Radu Coldea for sharing their data in Ref.\ \cite{Headings2010}. The experiment was performed at the ADRESS beamline of the Swiss Light Source using the SAXES instrument jointly built by Paul Scherrer Institut, Switzerland and Politecnico di Milano, Italy.  We acknowledge Vladimir Strocov for support at the ADRESS beamline. The work at Brookhaven is supported in part by the US DOE under contracts No.\ DEAC02-98CH10886 and MA-509-MACA and in part by the Center for Emergent Superconductivity, an Energy Frontier Research Center funded by the US DOE, Office of Basic Energy Sciences.

\section{Author contributions}
Experiment: M.P.M.D, J.P.H., R.S.S., C.M., K.J.Z., T.S.; sample growth: I.B.; sample characterization: I.B., J.P., R.S.S.\ and M.P.M.D; two-magnon calculations: B.D.P.\ and H.M.R.; data analysis and interpretation: M.P.M.D.\ J.P.H., J.v.d.B, T.S., C.M., K.J.Z.\ and H.M.R.. Project planning: J.P.H., M.P.M.D., T.S., I.B.; Paper writing: M.P.M.D and J.P.H., with contributions from all authors.
\section{Methods}

We preformed our RIXS experiments at the ADRESS beamline at the Swiss Light Source using the SAXES instrument. The total fluorescence yield at the Cu $L_3$ edge was measured at regular intervals and the incident energy was tuned to the peak in the absorption. We determined the combined energy resolution of the monochromator and spectrometer by measuring the elastic scattering from carbon tape, which was well described by a Gaussian function with a full width half maximum of 134~meV. The \mathbi{Q} resolution was better than 0.004~\AA{}$^{-1}$. The spectra were normalized to the intensity of the $dd$-excitations in order to account for differing amounts of \LCO{} probed in different films, and to facilitate the comparison of different films for a given scattering geometry. We denote the in-plane scattering vector, \mathbi{Q}, using the tetragonal \LCO{} unit cell $a=b=3.8$~\AA{}, with \mathbi{Q}= ($\pi$,0) parallel to the Cu-O-Cu bond direction.

For film synthesis, we employed a unique atomic layer-by-layer molecular-beam epitaxy system equipped with advanced tools for in-situ surface analysis including reflection high-energy-electron diffraction (RHEED) and time-of-flight ion-scattering spectroscopy. Using this technique, we reproducibly fabricate single-crystal films with atomically smooth surfaces and interfaces as well as heterostructures and superlattices with superconducting or insulating layers that can be down to one unit cell thick \cite{Gozar2008,Logvenov2009}. Digital layer-by-layer growth and the capability to maintain atomic-scale smoothness were both crucial for the presented study. The films were grown on single-crystal LaSrAlO$_4$ substrates, each with the 10$\times$10~mm$^{2}$ surface polished perpendicular to the (001) direction, under $9\times 10^{-6}$~Torr of ozone and at a substrate temperature of about 700~$^{\circ}$C. The deposition rates were measured by a quartz crystal oscillator before growth and controlled in real time using a custom-made atomic absorption spectroscopy system. The quality of the film growth was checked by monitoring RHEED intensity oscillations, which provide digital information on the film thickness. The films were subsequently annealed under high vacuum to drive out all the interstitial oxygen and avoid inadvertent oxygen doping. The sample characterization described in supplementary materials [URL to be inserted] shows that these films are a good realization of 1 and 2 unit cell thick \LCO{} layers and muon spin rotation measurement shows that isolated \LCO{} layers prepared in a similar way do not order \cite{Suter2011}.

The two magnon excitation spectrum $S^{zz}(\mathbi{Q},\omega)$ was calculated following Ref.\ \cite{Piazza2011} for a Hubbard model relevant to \LCO{}. The first, second, and third nearest-neighbor hopping parameters were $t=492$~meV, $t''=-207$~meV and $t'''=-45$~meV \cite{Piazza2011} as determined by fitting to neutron scattering measurements \cite{Headings2010}. The Coulomb replusion was fixed at $U=3.5$~eV .


\begin{thebibliography}{10}
\expandafter\ifx\csname url\endcsname\relax
  \def\url#1{\texttt{#1}}\fi
\expandafter\ifx\csname urlprefix\endcsname\relax\def\urlprefix{URL }\fi
\providecommand{\bibinfo}[2]{#2}
\providecommand{\eprint}[2][]{\url{#2}}

\bibitem{Hayden2004}
\bibinfo{author}{Hayden, S.~M.}, \bibinfo{author}{Mook, H.~A.},
  \bibinfo{author}{Dai, P.}, \bibinfo{author}{Perring, T.~G.} \&
  \bibinfo{author}{Dogan, F.}
\newblock \bibinfo{title}{The structure of the high-energy spin excitations in
  a high-transition-temperature superconductor}.
\newblock \emph{\bibinfo{journal}{Nature}} \textbf{\bibinfo{volume}{429}},
  \bibinfo{pages}{531--534} (\bibinfo{year}{2004}).

\bibitem{Vignolle2007}
\bibinfo{author}{Vignolle, B.} \emph{et~al.}
\newblock \bibinfo{title}{Two energy scales in the spin excitations of the
  high-temperature superconductor {La$_{2-x}$Sr$_x$CuO$_4$}}.
\newblock \emph{\bibinfo{journal}{Nature Phys.}} \textbf{\bibinfo{volume}{3}},
  \bibinfo{pages}{163--167} (\bibinfo{year}{2007}).

\bibitem{Li2010}
\bibinfo{author}{Li, Y.} \emph{et~al.}
\newblock \bibinfo{title}{Hidden magnetic excitation in the pseudogap phase of
  a high-{$T_c$} superconductor}.
\newblock \emph{\bibinfo{journal}{Nature}} \textbf{\bibinfo{volume}{468}},
  \bibinfo{pages}{283--285} (\bibinfo{year}{2010}).

\bibitem{LeTacon2011}
\bibinfo{author}{Le~Tacon, M.} \emph{et~al.}
\newblock \bibinfo{title}{Intense paramagnon excitations in a large family of
  high-temperature superconductors}.
\newblock \emph{\bibinfo{journal}{Nature Phys.}}
  \textbf{\bibinfo{volume}{advance online publication}}, \bibinfo{pages}{--}
  (\bibinfo{year}{2011}).
\newblock \urlprefix\url{http://dx.doi.org/10.1038/nphys2041}.

\bibitem{Coldea2001}
\bibinfo{author}{Coldea, R.} \emph{et~al.}
\newblock \bibinfo{title}{Spin waves and electronic interactions in
  {La$_2$CuO$_4$}}.
\newblock \emph{\bibinfo{journal}{Phys. Rev. Lett.}}
  \textbf{\bibinfo{volume}{86}}, \bibinfo{pages}{5377--5380}
  (\bibinfo{year}{2001}).

\bibitem{Headings2010}
\bibinfo{author}{Headings, N.~S.}, \bibinfo{author}{Hayden, S.~M.},
  \bibinfo{author}{Coldea, R.} \& \bibinfo{author}{Perring, T.~G.}
\newblock \bibinfo{title}{Anomalous high-energy spin excitations in the
  high-{$T_c$} superconductor-parent antiferromagnet {La$_2$CuO$_4$}}.
\newblock \emph{\bibinfo{journal}{Phys. Rev. Lett.}}
  \textbf{\bibinfo{volume}{105}}, \bibinfo{pages}{247001}
  (\bibinfo{year}{2010}).

\bibitem{Braicovich2010}
\bibinfo{author}{Braicovich, L.} \emph{et~al.}
\newblock \bibinfo{title}{Magnetic excitations and phase separation in the
  underdoped {La$_{2-x}$Sr$_x$CuO$_4$ } superconductor measured by resonant
  inelastic x-ray scattering}.
\newblock \emph{\bibinfo{journal}{Phys. Rev. Lett.}}
  \textbf{\bibinfo{volume}{104}}, \bibinfo{pages}{077002}
  (\bibinfo{year}{2010}).

\bibitem{Mermin1966}
\bibinfo{author}{Mermin, N.~D.} \& \bibinfo{author}{Wagner, H.}
\newblock \bibinfo{title}{Absence of ferromagnetism or antiferromagnetism in
  one- or two-dimensional isotropic {Heisenberg }models}.
\newblock \emph{\bibinfo{journal}{Phys. Rev. Lett.}}
  \textbf{\bibinfo{volume}{17}}, \bibinfo{pages}{1133} (\bibinfo{year}{1966}).

\bibitem{Sandvik2001}
\bibinfo{author}{Sandvik, A.} \& \bibinfo{author}{Singh, R.}
\newblock \bibinfo{title}{High-energy magnon dispersion and multimagnon
  continuum in the two-dimensional {Heisenberg} antiferromagnet}.
\newblock \emph{\bibinfo{journal}{Phys. Rev. Lett.}}
  \textbf{\bibinfo{volume}{86}}, \bibinfo{pages}{528--531}
  (\bibinfo{year}{2001}).

\bibitem{Ho2001}
\bibinfo{author}{Ho, C.-M.}, \bibinfo{author}{Muthukumar, V.},
  \bibinfo{author}{Ogata, M.} \& \bibinfo{author}{Anderson, P.}
\newblock \bibinfo{title}{Nature of spin excitations in two-dimensional {Mott}
  insulators: Undoped cuprates and other materials}.
\newblock \emph{\bibinfo{journal}{Phys. Rev. Lett.}}
  \textbf{\bibinfo{volume}{86}}, \bibinfo{pages}{1626--1629}
  (\bibinfo{year}{2001}).

\bibitem{Christensen2007}
\bibinfo{author}{Christensen, N.~B.} \emph{et~al.}
\newblock \bibinfo{title}{Quantum dynamics and entanglement of spins on a
  square lattice.}
\newblock \emph{\bibinfo{journal}{PNAS}} \textbf{\bibinfo{volume}{104}},
  \bibinfo{pages}{15264--9} (\bibinfo{year}{2007}).

\bibitem{Anderson1987}
\bibinfo{author}{Anderson, P.~W.}
\newblock \bibinfo{title}{The resonating valence bond state in {La$_2$CuO$_4$}
  and superconductivity}.
\newblock \emph{\bibinfo{journal}{Science}} \textbf{\bibinfo{volume}{235}},
  \bibinfo{pages}{1196--8} (\bibinfo{year}{1987}).

\bibitem{Hsu1990}
\bibinfo{author}{Hsu, T.~C.}
\newblock \bibinfo{title}{Spin waves in the flux-phase description of the s=1/2
  heisenberg antiferromagnet}.
\newblock \emph{\bibinfo{journal}{Phys. Rev. B}} \textbf{\bibinfo{volume}{41}},
  \bibinfo{pages}{11379--11387} (\bibinfo{year}{1990}).

\bibitem{Manousakis1991}
\bibinfo{author}{Manousakis, E.}
\newblock \bibinfo{title}{The spin-\textonehalf{} heisenberg antiferromagnet on
  a square lattice and its application to the cuprous oxides}.
\newblock \emph{\bibinfo{journal}{Rev. Mod. Phys.}}
  \textbf{\bibinfo{volume}{63}}, \bibinfo{pages}{1--62} (\bibinfo{year}{1991}).

\bibitem{Keimer1992}
\bibinfo{author}{Keimer, B.} \emph{et~al.}
\newblock \bibinfo{title}{Magnetic excitations in pure, lightly doped, and
  weakly metallic {La$_2$CuO$_4$}}.
\newblock \emph{\bibinfo{journal}{Phys. Rev. B}} \textbf{\bibinfo{volume}{46}},
  \bibinfo{pages}{14034--14053} (\bibinfo{year}{1992}).

\bibitem{Chakravarty1989}
\bibinfo{author}{Chakravarty, S.}, \bibinfo{author}{Halperin, B.} \&
  \bibinfo{author}{Nelson, D.}
\newblock \bibinfo{title}{{Two-dimensional quantum Heisenberg antiferromagnet
  at low temperatures}}.
\newblock \emph{\bibinfo{journal}{Phys. Rev. B}} \textbf{\bibinfo{volume}{39}},
  \bibinfo{pages}{2344--2371} (\bibinfo{year}{1989}).

\bibitem{Suter2011}
\bibinfo{author}{Suter, A.} \emph{et~al.}
\newblock \bibinfo{title}{Two-dimensional magnetic and superconducting phases
  in metal-insulator {La$_{2-x}$Sr$_x$CuO$_4$} superlattices measured by
  muon-spin rotation}.
\newblock \emph{\bibinfo{journal}{Phys. Rev. Lett.}}
  \textbf{\bibinfo{volume}{106}}, \bibinfo{pages}{237003}
  (\bibinfo{year}{2011}).

\bibitem{Tranquada2005}
\bibinfo{author}{Tranquada, J.~M.}
\newblock \emph{\bibinfo{title}{Treatise of High Temperature
  Superconductivity}}, chap. \bibinfo{chapter}{Neutron Scattering Studies of
  Antiferromagnetic Correlations in Cuprates} (\bibinfo{publisher}{Springer},
  \bibinfo{year}{2005}).

\bibitem{Schlappa2009}
\bibinfo{author}{Schlappa, J.} \emph{et~al.}
\newblock \bibinfo{title}{Collective magnetic excitations in the spin ladder
  {Sr$_{14}$Cu$_{24}$O$_{41}$} measured using high-resolution {Resonant
  Inelastic X-Ray Scattering}}.
\newblock \emph{\bibinfo{journal}{Phys. Rev. Lett.}}
  \textbf{\bibinfo{volume}{103}}, \bibinfo{pages}{047401}
  (\bibinfo{year}{2009}).

\bibitem{Guarise2010}
\bibinfo{author}{Guarise, M.} \emph{et~al.}
\newblock \bibinfo{title}{Measurement of magnetic excitations in the
  two-dimensional antiferromagnetic {Sr$_2$CuO$_2$Cl$_2$} insulator using
  resonant x-ray scattering: Evidence for extended interactions}.
\newblock \emph{\bibinfo{journal}{Phys. Rev. Lett.}}
  \textbf{\bibinfo{volume}{105}}, \bibinfo{pages}{157006}
  (\bibinfo{year}{2010}).

\bibitem{Ament2011}
\bibinfo{author}{Ament, L. J.~P.}, \bibinfo{author}{van Veenendaal, M.},
  \bibinfo{author}{Devereaux, T.~P.}, \bibinfo{author}{Hill, J.~P.} \&
  \bibinfo{author}{van~den Brink, J.}
\newblock \bibinfo{title}{Resonant inelastic x-ray scattering studies of
  elementary excitations}.
\newblock \emph{\bibinfo{journal}{Rev. Mod. Phys.}}  (\bibinfo{year}{2011}).

\bibitem{Piazza2011}
\bibinfo{author}{{Dalla Piazza}, B.} \emph{et~al.}
\newblock \bibinfo{title}{{Unified quantitative model for magnetic and
  electronic spectra of the undoped cuprates}}.
\newblock \emph{\bibinfo{journal}{ArXiv e-prints}}  (\bibinfo{year}{2011}).
\newblock \eprint{1104.4224}.

\bibitem{Ronnow2001}
\bibinfo{author}{R\o{}nnow, H.~M.} \emph{et~al.}
\newblock \bibinfo{title}{Spin dynamics of the 2d spin $\frac{1}{2}$ quantum
  antiferromagnet copper deuteroformate tetradeuterate (cftd)}.
\newblock \emph{\bibinfo{journal}{Phys. Rev. Lett.}}
  \textbf{\bibinfo{volume}{87}}, \bibinfo{pages}{037202}
  (\bibinfo{year}{2001}).

\bibitem{Singh1995}
\bibinfo{author}{Singh, R. R.~P.} \& \bibinfo{author}{Gelfand, M.~P.}
\newblock \bibinfo{title}{Spin-wave excitation spectra and spectral weights in
  square lattice antiferromagnets}.
\newblock \emph{\bibinfo{journal}{Phys. Rev. B}} \textbf{\bibinfo{volume}{52}},
  \bibinfo{pages}{R15695--R15698} (\bibinfo{year}{1995}).

\bibitem{Ament3mag}
\bibinfo{author}{{Ament}, L.~J.~P.} \& \bibinfo{author}{{van den Brink}, J.}
\newblock \bibinfo{title}{{Strong Three-magnon Scattering in Cuprates by
  Resonant X-rays}}.
\newblock \emph{\bibinfo{journal}{ArXiv e-prints}}  (\bibinfo{year}{2010}).
\newblock \eprint{1002.3773}.

\bibitem{Igarashi2011}
\bibinfo{author}{{Igarashi}, J.-I.} \& \bibinfo{author}{{Nagao}, T.}
\newblock \bibinfo{title}{{Magnetic excitations in $L$-edge resonant inelastic
  x-ray scattering from cuprate compounds}}.
\newblock \emph{\bibinfo{journal}{ArXiv e-prints}}  (\bibinfo{year}{2011}).
\newblock \eprint{1104.4683}.

\bibitem{Ament2009}
\bibinfo{author}{Ament, L.}, \bibinfo{author}{Ghiringhelli, G.},
  \bibinfo{author}{Sala, M.}, \bibinfo{author}{Braicovich, L.} \&
  \bibinfo{author}{van~den Brink, J.}
\newblock \bibinfo{title}{Theoretical demonstration of how the dispersion of
  magnetic excitations in cuprate compounds can be determined using resonant
  inelastic x-ray scattering}.
\newblock \emph{\bibinfo{journal}{Phys. Rev. Lett.}}
  \textbf{\bibinfo{volume}{103}}, \bibinfo{pages}{117003}
  (\bibinfo{year}{2009}).

\bibitem{Haverkort2010}
\bibinfo{author}{Haverkort, M.~W.}
\newblock \bibinfo{title}{Theory of resonant inelastic x-ray scattering by
  collective magnetic excitations}.
\newblock \emph{\bibinfo{journal}{Phys. Rev. Lett.}}
  \textbf{\bibinfo{volume}{105}}, \bibinfo{pages}{167404}
  (\bibinfo{year}{2010}).

\bibitem{Huberman2005}
\bibinfo{author}{Huberman, T.} \emph{et~al.}
\newblock \bibinfo{title}{{Two-magnon excitations observed by neutron
  scattering in the two-dimensional spin-$\frac{5}{2}$ Heisenberg
  antiferromagnet Rb$_2$MnF$_4$}}.
\newblock \emph{\bibinfo{journal}{Phys. Rev. B}} \textbf{\bibinfo{volume}{72}}
  (\bibinfo{year}{2005}).

\bibitem{Gozar2008}
\bibinfo{author}{Gozar, A.} \emph{et~al.}
\newblock \bibinfo{title}{High-temperature interface superconductivity between
  metallic and insulating copper oxides.}
\newblock \emph{\bibinfo{journal}{Nature}} \textbf{\bibinfo{volume}{455}},
  \bibinfo{pages}{782--5} (\bibinfo{year}{2008}).

\bibitem{Logvenov2009}
\bibinfo{author}{Logvenov, G.}, \bibinfo{author}{Gozar, A.} \&
  \bibinfo{author}{Bo\v{z}ovi\'{c}, I.}
\newblock \bibinfo{title}{High-temperature superconductivity in a single
  copper-oxygen plane}.
\newblock \emph{\bibinfo{journal}{Science}} \textbf{\bibinfo{volume}{326}},
  \bibinfo{pages}{699--702} (\bibinfo{year}{2009}).

\end{thebibliography}
\end{document}